# Accurate typhoon intensity forecasts using a non-iterative spatiotemporal transformer model

Hongyu Qu[1], Hongxiong Xu[2*], Lin Dong[1], Chunyi Xiang[1], Gaozhen Nie[1]

[1]National Meteorological Centre, Beijing, 100081, China

[2]State Key Laboratory of Severe Weather, Chinese Academy of Meteorological Sciences, China Meteorological Administration, Beijing, 100081, China

August 24, 2025
Dateline

---

Corresponding author:   Dr. Hongxiong Xu
State Key Laboratory of Severe Weather
Meteorological Science and Technology
Chinese Academy of Meteorological Sciences
No. 46, Zhongguancun South Street, Haidian District
Beijing, P. R. China, 100081
Email: xuhx@cma.gov.cn

**ABSTRACT**

Accurate forecasting of tropical cyclone (TC) intensity—particularly during periods of rapid intensification and rapid weakening—remains a challenge for operational meteorology, with high-stakes implications for disaster preparedness and infrastructure resilience. Recent advances in machine learning have yielded notable progress in TC prediction; however, most existing systems provide forecasts that degrade rapidly in extreme regimes and lack long-range consistency. Here we introduce TIFNet, a transformer-based forecasting model that generates non-iterative, 5-day intensity trajectories by integrating high-resolution global forecasts with a historical-evolution fusion mechanism. Trained on reanalysis data and fine-tuned with operational data, TIFNet consistently outperforms operational numerical models across all forecast horizons, delivering robust improvements across weak, strong, and super typhoon categories. In rapid intensity change regimes—long regarded as the most difficult to forecast—TIFNet reduces forecast error by 29–43% relative to current operational baselines. These results represent a substantial advance in artificial-intelligence-based TC intensity forecasting, especially under extreme conditions where traditional models consistently underperform.

**INTRODUCTION**

Tropical cyclones (TCs) are among the most devastating natural hazards, responsible for widespread loss of life, large-scale infrastructure damage, and severe economic disruption worldwide[1,2]. Their multifaceted threats—including destructive winds, torrential rainfall, flooding, and storm surges—pose particular risks to densely populated coastal regions[3]. For example, the western North Pacific (WNP) basin—the most active TC region globally—is fringed by vulnerable megacities and economies, making the region particularly susceptible to catastrophic impacts[4,5]. Hence, accurate forecasts of TC intensity, especially during periods of rapid intensification (RI) and rapid weakening (RW), remain essential for effective disaster preparedness and timely mitigation of TC impacts[6].

Despite decades of progress, prediction of TC intensity continues to lag substantially behind TC track forecasting[7]. This gap reflects the inherent limitations of the two prevailing approaches: dynamical and statistical models. Dynamical models explicitly integrate the governing equations of atmospheric motion and thermodynamics, assimilating vast observational datasets to represent the coupled atmosphere–ocean system[8,9]. Although advances in computing and data assimilation have markedly reduced TC track forecast errors, improvements in TC intensity forecasts have been far more limited[10,11]. Statistical models, by contrast, are built on empirical relationships between storm intensity change and large-scale environmental predictors derived from historical cases[12]. While computationally efficient, their reliance on historically derived patterns limits their ability to capture rare, nonlinear events such as RI and RW[13,14]. Consequently, both approaches fall short in representing the strongly nonlinear, multiscale processes that govern TC evolution[15].

The rapid evolution of artificial intelligence (AI) has provided a transformative alternative approach to TC intensity forecasting[16]. Deep learning (DL) offers a powerful alternative by representing complex, nonlinear interactions across multiple scales that have long eluded dynamical and statistical approaches[17]. Early studies applied recurrent neural network or improved versions such as long short-term memory (LSTM) to forecast intensity from historical time series[18,19]. While these approaches demonstrated the potential of DL, their reliance on simple temporal features proved insufficient to capture the evolving state of TCs and their relationship with the surrounding environmental fields; consequently, improvements in forecast performance have remained limited. Building on this, more recent efforts incorporated 3D spatial datasets—most notably the ERA5 dataset—to represent both spatial and temporal variability[20–26]. By fusing 3D environmental predictors with advanced neural architectures, these studies achieved meaningful gains, particularly in short-range forecasts up to 24 h. The latest generation of DL models further integrates multisource environmental and satellite data with attention mechanisms, convolutional networks, and transformer encoders, delivering substantial improvements in the representation of TC structure and intensity evolution[26,27].

Collectively, these advances highlight that integrating higher-dimensional datasets with advanced neural architectures provides a promising pathway for improving the skill and reliability of TC intensity forecasts. Nevertheless, two critical gaps remain. First, most existing systems depend heavily on initial atmospheric states or short sequences of past observations, leading to sharp decline in forecast skill beyond 24 h and further deterioration at longer ranges as storm–environment feedbacks amplify uncertainty[19,21–23,28,29]. Although incorporation of physical constraints has shown potential to alleviate this limitation[30], purely data-driven models still perform poorly at extended lead times. Second, these models systematically underpredict rapid changes in intensity because the scarcity of extreme events biases learning toward moderate trajectories. Thus, they often fail to capture the full evolution of TC intensity, further suppressing extreme transitions and leaving RI and RW regimes poorly represented[31]. Together, these limitations—loss of skill at extended lead times and failure to capture rapid transitions—represent the central barriers to advancing TC intensity forecasting.

To address these challenges, we developed the Tropical cyclone Intensity Forecasting Network (TIFNet)—a spatiotemporal transformer encoder–decoder that integrates high-resolution environmental features from state-of-the-art global forecasts with a historical-evolution fusion mechanism. Departing from the iterative time-stepping paradigm of numerical models, TIFNet generates coherent 0–120 h intensity trajectories in a single forward pass. This design fully exploits 4D spatiotemporal global forecast fields, enhancing temporal consistency, reducing cumulative errors, and improving robustness to initialization biases.

**RESULTS**

**TIFNet Model.**

TIFNet is a DL model for forecasting TC intensity that is based on a transformer encoder–decoder architecture[32]. The methodology involves training a transformer-based encoder–decoder network to take atmospheric fields and storm history as input, and then predict future intensity trajectories as output. The model was first pretrained on ERA5 data spanning 1990–2020 to leverage long-term historical variability, and subsequently fine-tuned with high-resolution forecasts from the ECMWF's Integrated Forecasting System (IFS) covering 2020–2021 to ensure consistency with real-time applications. For operational forecasting, the model ingests IFS analyses and forecasts at initialization and then outputs maximum sustained wind speed at 6-h intervals for lead times of up to 120 h. Each input sample consists of two temporal segments: the pre-initialization segment that provides 24 h of multivariate atmospheric analyses and cyclone intensity evolution, and the post-initialization segment that provides forecasted environmental fields for the subsequent 120 h.

The model architecture is illustrated in Fig. 1. Atmospheric variables for both the encoder and the decoder are initially projected into spatiotemporal patches via 3D convolution for upper-air fields and 2D convolution for surface fields. The encoder processes the pre-initialization fields and the cyclone history, while the decoder processes the post-initialization forecast fields. Temporal and spatial embeddings are incorporated at each stage. The encoded and decoded representations are fused via a cross-attention mechanism, allowing future-state queries to attend to past dynamics. The fused tokens are subsequently passed through a multi-head self-attention mechanism and a parameter-shared feedforward neural network to predict intensity changes ($\Delta Vmax$) at each forecast step.

Unlike conventional numerical weather models[8,33] and many AI-based forecasting systems[16,26,28,29,34,35], TIFNet is non-iterative, i.e., it generates the entire 0–120-h trajectory in a single forward pass. This design mitigates cumulative error propagation, improves temporal consistency, and enables skillful forecasts at extended lead times. By integrating high-resolution global fields across multiple vertical levels, TIFNet learns multiscale spatiotemporal patterns with robustness, achieving marked improvements in RI and RW regimes—scenarios where existing systems typically falter.

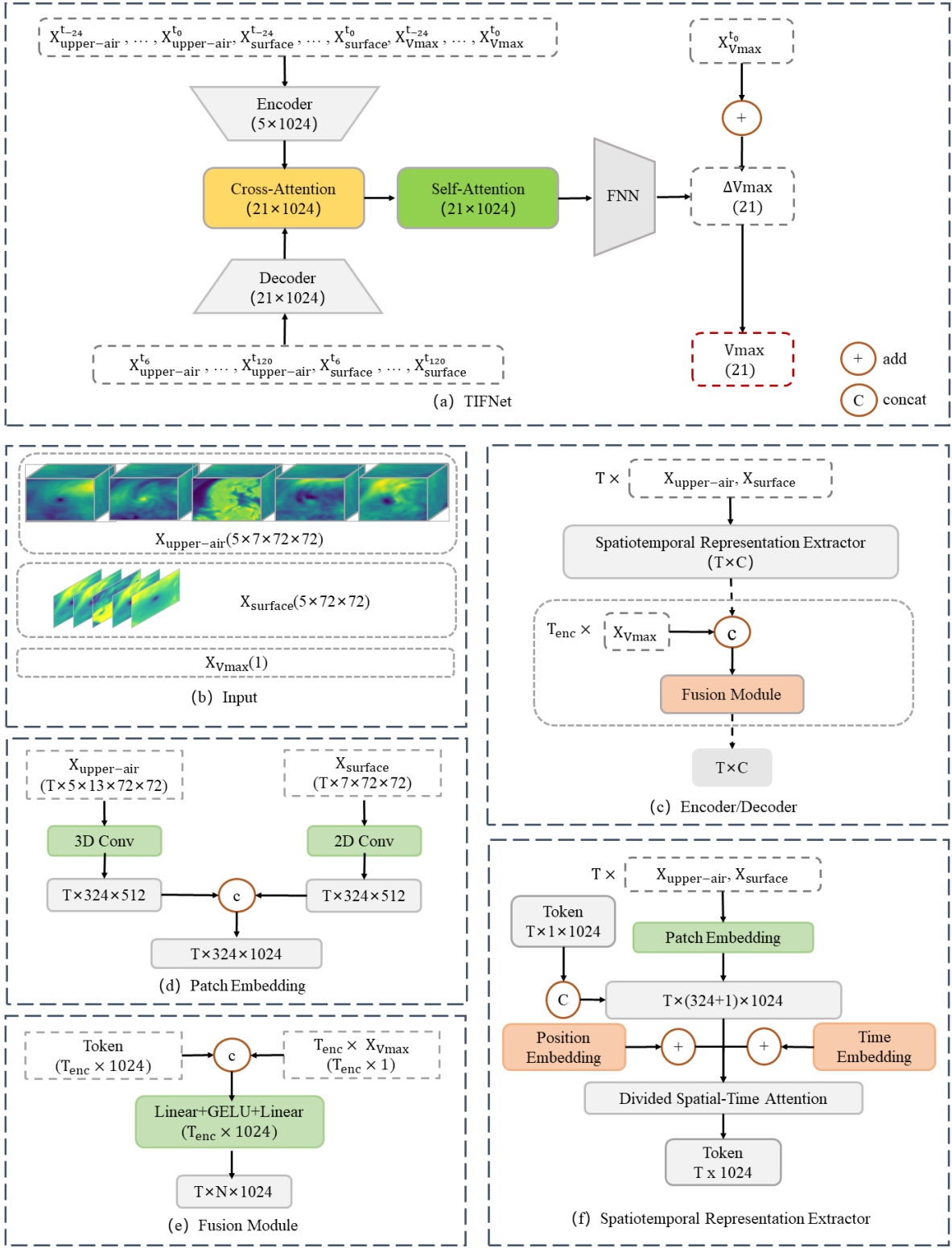

**Fig. 1 | Architecture of the Tropical cyclone Intensity Forecasting Network (TIFNet).**

(a) Schematic overview of the encoder–decoder framework, in which future meteorological forecast fields are fused with prior fields and the historical evolution of tropical cyclone intensity via a cross-attention mechanism, and then passed to a feedforward neural network (FNN) to predict future maximum wind speed (Vmax). (b) Model inputs include upper-air fields at seven pressure levels, surface variables, and historical intensity records. (c) Encoder and decoder structure featuring a spatiotemporal representation extractor and a fusion module, the latter of which is currently implemented solely within the encoder. (d) Patch embedding of atmospheric inputs using 3D convolution for upper-air fields and 2D convolution for surface variables. (e) Fusion module that combines environmental features with cyclone intensity history. (f) Spatiotemporal representation extractor incorporating patch embedding, positional and temporal encoding, and divided spatial-time attention mechanism.

**Experimental setup.**

We assessed TIFNet's forecasting skill through retrospective experiments on independent cases from the 2023-2024 typhoon seasons. All forecasts were initialized every 6-h using ECMWF IFS environmental fields together with best-track intensity records from the China Meteorological Administration (CMA). Performance was assessed using the mean absolute error (MAE) in the 10-m maximum sustained wind speed at each forecast lead time, following established best-track verification protocols. Forecast trajectories were generated over a 120-h horizon at 6-h intervals, and results were stratified by cyclone intensity, intensity-change regime, and forecast range.

TIFNet was evaluated against three operational numerical models: the ECMWF IFS, widely regarded as the global "gold standard" in TC forecasting; the NCEP GFS, one of the most widely distributed and operationally relied-upon global systems; and the CMA TYM, a regional typhoon model used extensively in China and optimized for prediction of TC intensity in the WNP basin. Each model was evaluated using the same initialization times and best-track references. This combination of global and regional baselines provides a rigorous and representative benchmark for assessing TIFNet's skill.

Identical forecast records were extracted across all models, yielding 41 TCs across the WNP basin—encompassing two complete typhoon seasons. The dataset captured wide diversity in meteorological scenarios—including open-ocean cyclones, recurring storms, and multiple landfalling events—and encompassed the full intensity range from tropical storm (TS) to super typhoon (SuperTY) (a summary of TC intensity categories is presented in Table 1). Critically, it also included RI and RW episodes, alongside numerous periods of gradual change in intensity (Fig. 2a), ensuring that both typical and extreme forecast regimes were represented.

**Table 1 | Intensity categories of tropical cyclones**[36]

| Category | Maximum wind speed(m s$^{-1}$) |
|---|---|
| Tropical depression(TD) | 10.8 – 17.1 |
| Tropical storm(TS) | 17.2 – 24.4 |
| Severe tropical storm(STS) | 24.5 – 32.6 |
| Typhoon(TY) | 32.7 – 41.4 |
| Severe typhoon(STY) | 41.5 – 50.9 |
| Super typhoon(SuperTY) | ⩾ 51.0 |

**Comparative assessment of TIFNet and operational forecast systems**

Fig. 2b summarizes the comparative forecast skill. It is evident that TIFNet consistently outperforms all numerical models across all lead times. At 12 h, TIFNet reduces MAE by 72% relative to the IFS and by more than 48% compared with both the GFS and the TYM. This advantage persists through the 48-h horizon—where most ML-based systems typically plateau[19,21–23,29]—with corresponding MAE reductions of 50% (IFS), 28% (GFS), and 32% (TYM), respectively.

Importantly, TIFNet maintains a clear lead at extended lead times. At 120 h, it achieves MAE reductions of 27% versus the IFS, 18% versus the TYM, and 6% versus the GFS. Although forecast errors increase with lead time across all models, TIFNet's early-cycle advantage enables it to sustain a meaningful margin—demonstrating that transformer-based architectures, when trained on reanalysis and driven by real-time forecast inputs, can retain skill beyond the conventional 48–72 h threshold for AI-based intensity prediction.

This extended-range skill is largely enabled by the inclusion of the IFS forecast fields.

Sensitivity experiments in which forecast fields were substituted with standard Gaussian noise exhibited a rapid increase in error growth beyond 48-h, closely replicating the degradation patterns observed in previous ML-based systems (Extended Data Fig. 1). Conversely, additional experimentation substituting the IFS forecasts with corresponding analyses throughout the forecast period substantially mitigated long-range error growth (Extended Data Fig. 2), indicating that performance limitations at longer lead times stem primarily from biases in the IFS forecasts—particularly in the representation of the large-scale circulation—rather than from deficiencies in either the TIFNet model architecture or the training procedure.

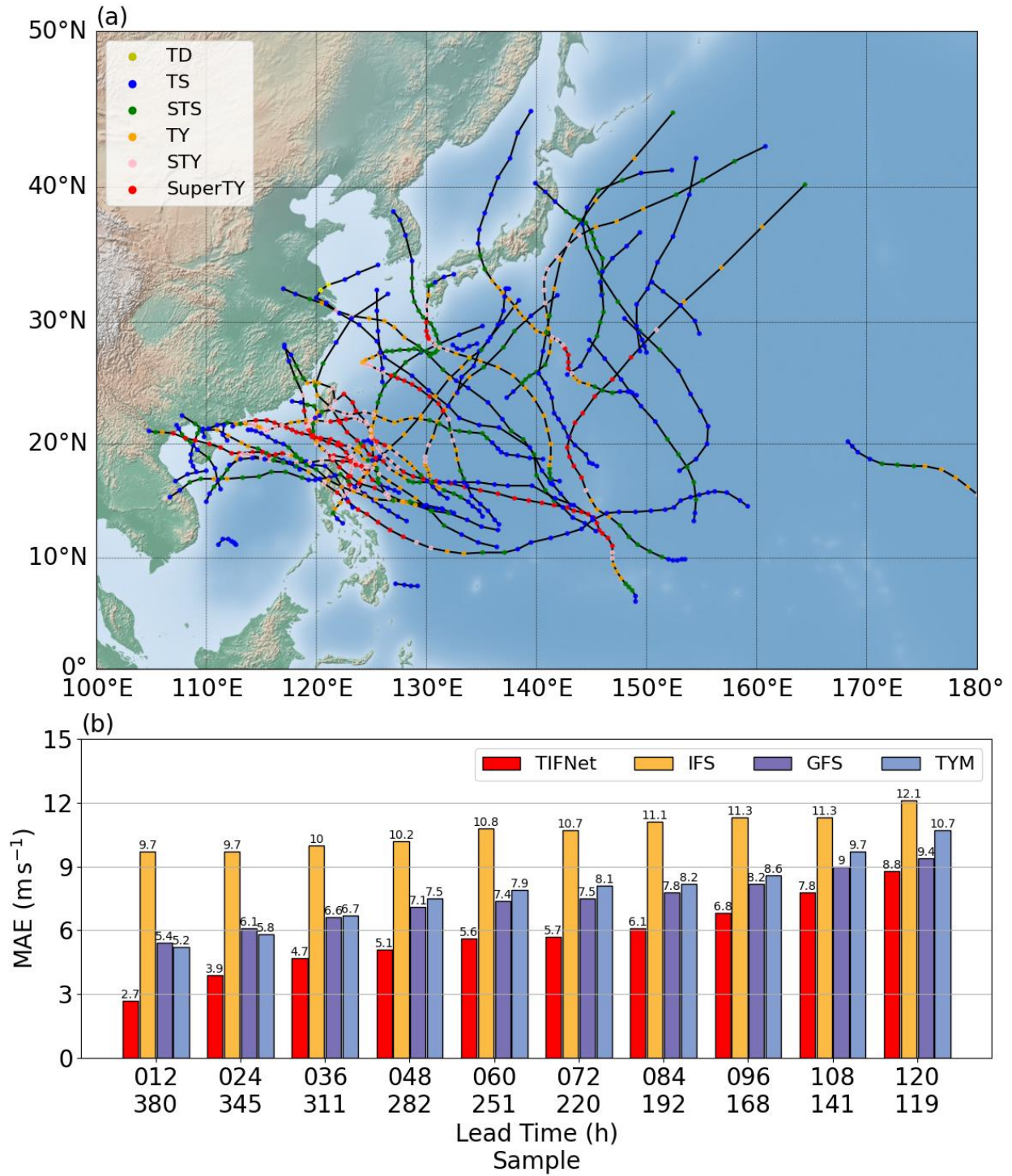

**Fig. 2 | TIFNet outperforms operational models in typhoon intensity forecasts over the western North Pacific basin.**

(a) Tropical cyclone tracks from the CMA best track dataset during the 2023–2024 seasons, color-

coded by intensity category (TD, TS, STS, TY, STY, and SuperTY). (b) Forecast performance in terms of mean absolute error (MAE; m s$^{-1}$) of maximum sustained wind speed at lead times of 12–120 h for TIFNet (red), the ECMWF IFS (yellow), the NCEP GFS (purple), and the CMA TYM (blue). Forecasts were evaluated against best-track intensity using identical initialization times. Bars show MAE averaged over all cases at each lead time, with values labeled above and corresponding sample sizes shown below the x-axis.

**Skillful forecast performance across intensity categories**

Building on the overall skill assessment, we further evaluated forecast performance across different TC intensity categories, as defined in Table 1. The MAE was calculated for each category and forecast window (Fig. 3). A clear trend emerged: TIFNet delivers competitive or superior performance across all intensity categories, with its advantage becoming more pronounced as storm intensity increases.

For TSs (Fig. 3a) and STSs (Fig. 3b), TIFNet achieves the lowest MAEs within the short-range window (⩽36 h). In both categories, it outperforms leading operational models, particularly at early lead times, although the gap narrows slightly beyond 36 h owing to underestimation tendencies in ECMWF IFS.

The advantage of TIFNet becomes more pronounced with stronger systems. For TYs (Fig. 3c), it maintains a consistent lead across all forecast horizons, while for STYs (Fig. 3d), the performance gap widens markedly. Notably, TIFNet demonstrates strong skill during SuperTY events (Fig. 3e)—a known weakness in many AI-based systems[37]. While previous models tend to underestimate peak intensities and fail to capture steep growth phases, TIFNet maintains substantially lower MAEs throughout, including at 120-h lead times.

Across all categories, forecast errors naturally increase with storm intensity across all models. However, TIFNet's architecture—combining spatiotemporal feature extraction with historical-evolution fusion—exhibits remarkable resilience under high-intensity conditions, particularly in the presence of SuperTY events. It not only rivals or outperforms numerical models in relation to weaker systems but also achieves unprecedented forecasting skill in terms of the most extreme and destructive regimes.

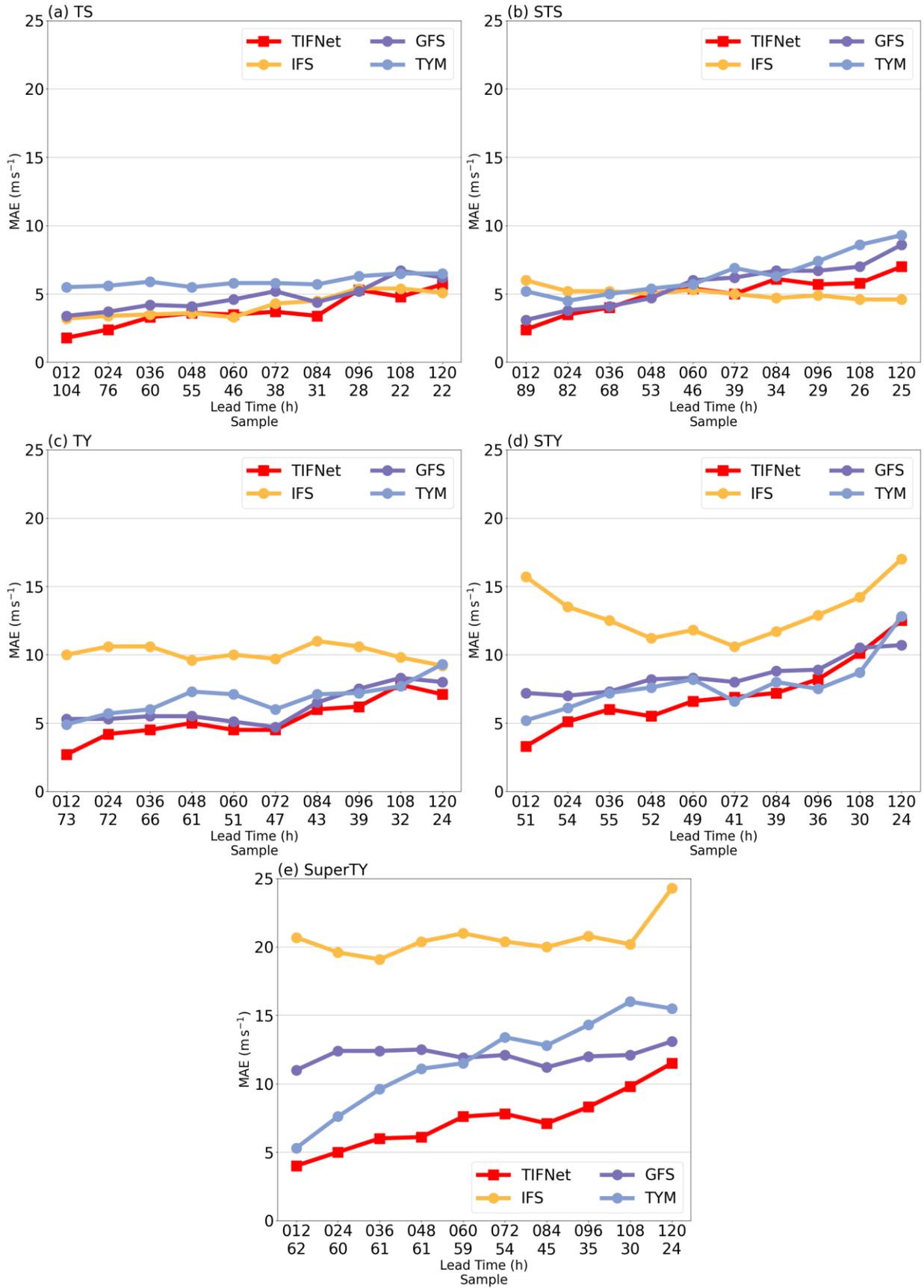

**Fig. 3 | Forecast error by TC intensity category.**

Mean absolute error (MAE; m s$^{-1}$) of intensity forecasts from 12 to 120 h, stratified by verifying intensity category: (a) TS, (b) STS, (c) TY, (d) STY, and (e) SuperTY. Forecast performance is compared across models: TIFNet (red), the ECMWF IFS (yellow), the NCEP GFS (purple), and the CMA TYM (blue). Sample sizes at each forecast lead time are indicated below the x-axis.

**Skill retention in regimes of rapid intensity change.**

Generally, RW and RI remain among the most difficult aspects of TC intensity prediction owing to their high dynamical complexity, short temporal scale, and intrinsic low predictability. Following the operational definition adopted by the World Meteorological Organization Typhoon Committee and widely used in recent intensity change studies, in accordance with the 24-h change in maximum wind speed, forecasts were classified into four categories: RW ($\leqslant -15$ m s$^{-1}$), general weakening ($-15$ to 0 m s$^{-1}$), general intensification (0 to 15 m s$^{-1}$), and RI ($\geqslant 15$ m s$^{-1}$)[13,38,39].

Fig. 4 shows the MAE stratified across these four regimes. For gradual intensity changes—both general weakening and general intensification—all models produce relatively low errors, yet TIFNet consistently delivers the most accurate forecasts. In general weakening cases, its MAE is 2.9 m s$^{-1}$, outperforming the IFS (6.5 m s$^{-1}$), the GFS (4.4 m s$^{-1}$), and the TYM (4.8 m s$^{-1}$). In general intensification cases, TIFNet achieves an MAE of 3.6 m s$^{-1}$, again lower than that of the IFS (5.2 m s$^{-1}$), the GFS (4.9 m s$^{-1}$), and the TYM (5.3 m s$^{-1}$).

More substantial differences emerge under rapid intensity transitions. In RW events—where maximum sustained winds drop by more than 15 m s$^{-1}$ in 24 h—forecast error increases sharply across all models but TIFNet's remains markedly lower at 4.7 m s$^{-1}$, compared with 12.6 m s$^{-1}$ for the IFS, 7.6 m s$^{-1}$ for the GFS, and 5.9 m s$^{-1}$ for the TYM. This corresponds to error reductions of 63% relative to the IFS, 38% relative to the GFS, and 20% relative to the TYM—highlighting TIFNet's ability to capture sharp post-peak decline that is often misrepresented by both global and regional dynamical models.

The most pronounced gains occur in RI regimes, a known weak spot for ML approaches owing to their rarity in the training data and complex coupling dynamics [40,41]. In RI cases, TIFNet achieves an MAE of 7.5 m s$^{-1}$, markedly better than that of the IFS (13.2 m s$^{-1}$), the GFS (10.5 m s$^{-1}$), and the TYM (11.0 m s$^{-1}$)—translating to error reductions of 43%, 29%, and 32%, respectively. Crucially, TIFNet captures not only the onset of RI but also the steep growth trajectories defining the most destructive SuperTYs—capabilities crucial for timely disaster preparedness and mitigation.

Collectively, these results demonstrate that TIFNet retains skill across all intensity-change regimes, with the largest advantages emerging in RI and RW scenarios—conditions under which conventional AI and dynamical systems typically degrade. The model's ability to generalize in such extreme, data-sparse conditions is likely attributable to its non-iterative architecture and historical-evolution fusion strategy (cf. Fig. 4 and Extended Data Fig. 3), thereby offering a meaningful step forward in addressing one of the longest-standing challenges in TC forecasting.

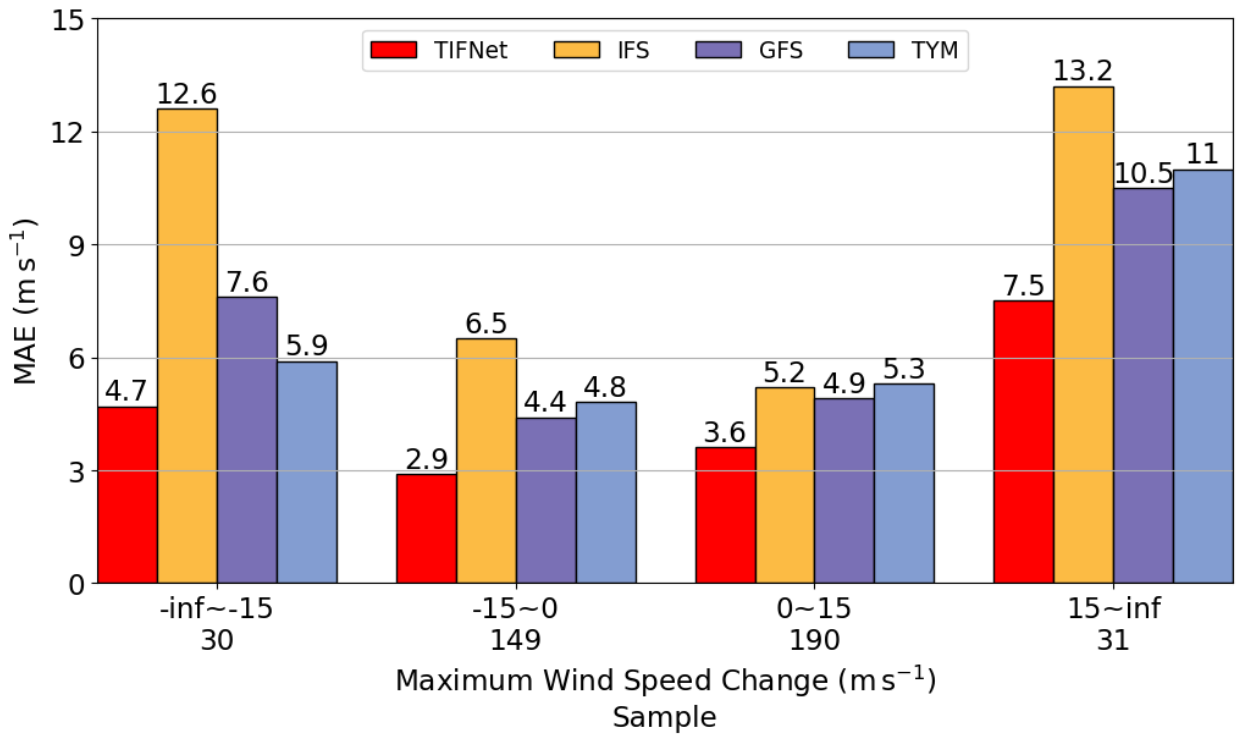

**Fig. 4 | Forecast errors stratified by intensity-change regime.**

Mean absolute error (MAE; m s$^{-1}$) for TIFNet (red), the ECMWF IFS (yellow), the NCEP GFS (purple), and the CMA TYM (blue), grouped by maximum 24-h wind-speed change (ΔVmax$_{24}$, m s$^{-1}$): $\Delta Vmax_{24} \leqslant -15$, $-15 < \Delta Vmax_{24} < 0$, $0 \leqslant \Delta Vmax_{24} < 15$, and $\Delta Vmax_{24} \geqslant 15$. MAE values are shown above each bar, with corresponding sample sizes indicated below the x-axis.

**Case-based evaluation in high-impact typhoon events**

To evaluate performance in real-world high-impact scenarios, we conducted detailed case analyses for three representative typhoons: DOKSURI (2305), BOLAVEN (2315), and YAGI (2411), spanning different tracks, intensity evolutions, and environmental contexts. Among these, DOKSURI was selected as the primary example because of its exceptional complexity, featuring two distinct RI events, multiple landfalls, and rapid post-landfall decay (Fig. 5a).

TIFNet reproduced this complex evolution with remarkable fidelity (total MAE: 4.0 m s$^{-1}$, Fig. 5b), capturing both RI events, the peak magnitudes, and the decay rates within a narrow error margin from the earliest forecast cycles. In contrast, the IFS (total MAE: 12.9 m s$^{-1}$, Fig. 5c) failed to capture the first RI and underestimated both peaks; the GFS (total MAE: 8.9 m s$^{-1}$, Fig. 5d) and the TYM (total MAE: 7.4 m s$^{-1}$, Fig. 5e) both exhibited systematic underprediction during intensification and initiated weakening too early. These results highlight TIFNet's capacity to sustain high forecast skill through rapid transitions where both global and regional operational models degrade sharply.

This advantage is further supported by results from BOLAVEN and YAGI (Extended Data Figs. 4 and 5), which follow a similar pattern: TIFNet consistently outperforms all baselines in capturing both the magnitude and the timing of peak intensities, especially during steep RI phases. Notably, this capability is closely tied to TIFNet's non-iterative prediction design, which allows the model to generate temporally coherent 5-day intensity trajectories in a single forward pass. In contrast, iterative variants of TIFNet exhibit notably smoother outputs that underrepresent the sharp gradients typical of RI and super typhoon events (Extended Data Figs. 3 and 6), leading to notable degradation in peak-intensity prediction. These results further highlight the importance of non-iterative architectures for forecasting extreme, rapidly evolving phenomena, where accumulated stepwise errors and over-smoothing in iterative approaches can severely limit predictive fidelity.

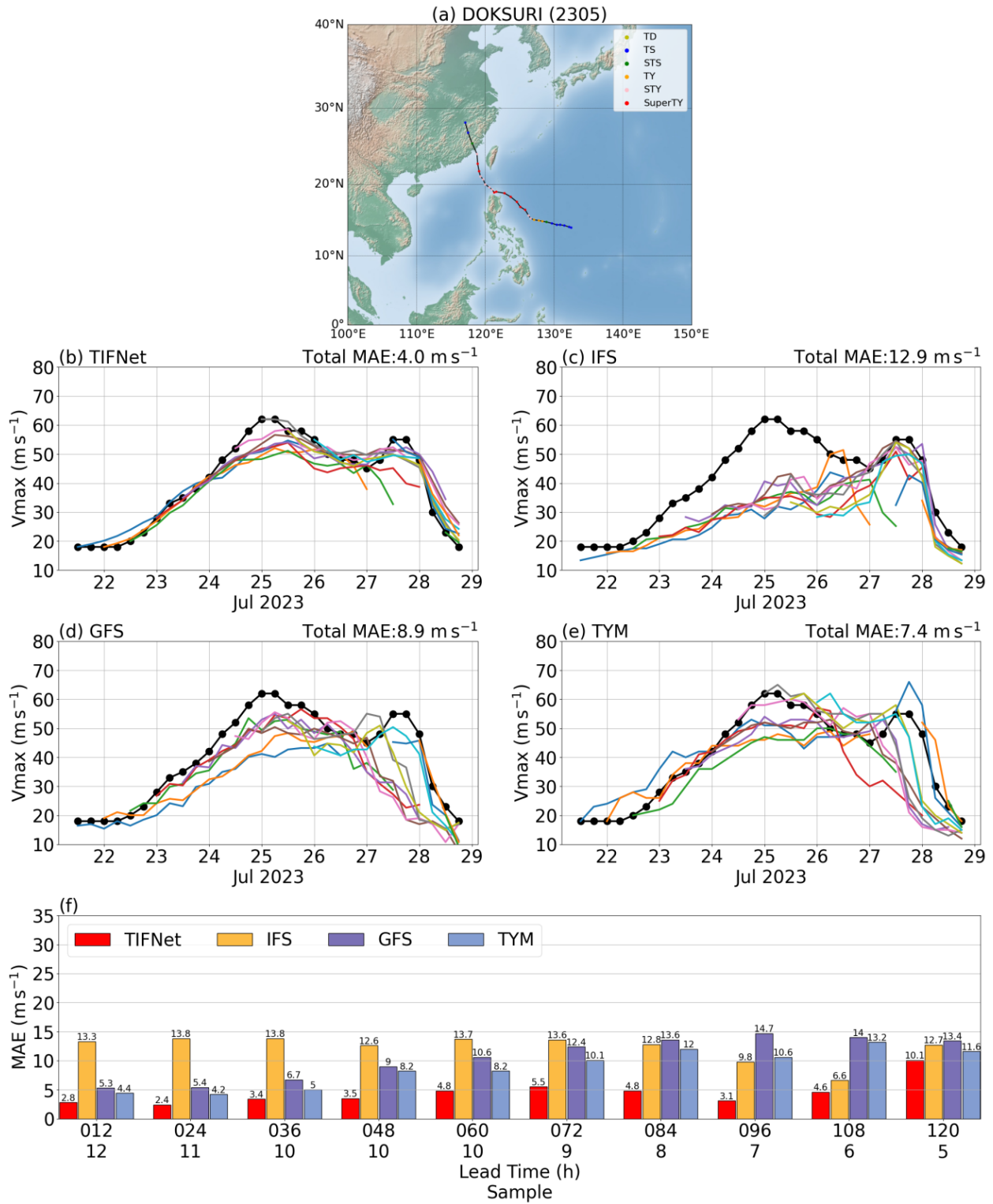

**Fig. 5 | Forecast validation for Typhoon DOKSURI (2305).**

(a) Tropical cyclone tracks from the CMA best track, color-coded by intensity category. (b–e) Intensity forecasts from TIFNet, the ECMWF IFS, the NCEP GFS and the CMA TYM (colored lines) compared with best-track observations (black) over successive forecast cycles. MAE (m s$^{-1}$) indicates the average forecast error across all cycles. (f) Lead-time-dependent MAE (12–120-h) for each model; sample sizes are shown below the x-axis.

**DISCUSSION**

This study introduced TIFNet, a transformer-based DL framework for TC intensity forecasting, which demonstrates superior forecast skill over operational numerical models across cyclone intensities, regimes of rapid change, and forecast horizons. The model's ability to generate coherent 5-day intensity trajectories in a single non-iterative pass marks a departure from both traditional stepwise numerical integration and common iterative ML approaches. Unlike many existing AI-based forecasting systems that smooth extreme transitions, TIFNet preserves resolution during RI and RW phases. Leveraging high-resolution global forecast fields and historical cyclone evolution, TIFNet consistently improves upon the performance of the ECMWF IFS, the NCEP GFS, and the CMA TYM, particularly in high-impact cases such as RI and super typhoons.

Despite these advances, several limitations remain. First, TIFNet's long-range skill is partly constrained by errors inherited from upstream numerical forecasts. Sensitivity experiments suggest that large-scale circulation biases in the driving IFS fields markedly degrade performance beyond 48 h (Extended Data Fig. 1). Incorporating ensemble forecasts or bias-corrected inputs could help mitigate these issues. Second, while TIFNet is effective in RI and RW regimes, further improvements might be achieved by augmenting the training data with synthetic or targeted extreme cases to better capture low-frequency events.

Looking forward, the architecture underpinning TIFNet—based on spatiotemporal attention and environmental–historical fusion—offers a scalable foundation for broader geophysical applications. Its capacity to learn structured multivariate dependencies across time and space from 4D data suggests potential generalization to other rapidly evolving hazards, such as extratropical cyclones, mesoscale convective systems, and flash flood–inducing events. Moreover, its core design principles could inform next-generation global AI weather models, improving spatial resolution, temporal coherence, and extreme-event predictability. In this context, models such as TIFNet can help close the gap between short-range precision and long-range resilience. As the frequency and intensity of extreme TCs increase under anthropogenic climate forcing, such capabilities are essential for advancing disaster preparedness and enhancing climate-resilient forecasting systems.

## METHODS

### Datasets

To train and evaluate our typhoon intensity forecasting network (TIFNet), we integrated three complementary data sources designed to capture both the historical context and real-time environmental drivers of tropical cyclones (TCs).

**CMA Best Track Dataset**[42]**.** Provided by the China Meteorological Administration, this dataset covers six-hourly records for tropical cyclone positions and intensities over the western North Pacific basin since 1949 and is widely used in typhoon research. Each record includes the 2-min maximum sustained wind at the cyclone center and minimum sea level pressure. We selected maximum sustained wind speed as the prediction target and incorporated historical intensity sequences preceding forecast initialization as model inputs to enhance performance.

**ERA5**[43]. The fifth-generation European Centre for Medium-Range Weather Forecasts reanalysis offers hourly global atmospheric fields from 1940 onward at approximately 31 km horizontal resolution, spanning up to 137 vertical levels. Consistent with established practice[23,28,34], we extracted zonal and meridional winds, temperature, geopotential height, and relative humidity across seven standard pressure levels (100, 200, 300, 500, 700, 850 and 950 hPa), together with sea surface temperature and six surface-level variables: 10-meter u wind component, 10-meter v wind component, sea-level pressure, 2 m temperature, topographic elevation, and land–sea mask. ERA5 data from 1990–2020 were used for model pretraining, while 2021–2022 served as validation.

**ECMWF IFS Forecasts**[44]**.** The ERA5 dataset lacks real-time availability and therefore we used IFS forecasts for operational-stage inputs. These forecasts provide surface-level variables at 0.125° resolution and upper-air variables at 0.25° resolution with a lead time up to 240 h. For alignment with ERA5, we bilinearly interpolated forecasts to 0.25°. Fine-tuning was conducted using ECMWF forecasts from 2000–2021, validation used data from 2022, and real-time evaluation employed forecasts from 2023–2024.

**Spatial Subsetting.** The ERA5 Reanalysis and the IFS forecast datasets were cropped to a spatial window centered on each cyclone. Although use of a 10° × 10° window is common[23,45], we used a broader 18° × 18° window to better capture environmental interactions while maintaining computational efficiency. Cyclone centers were identified via a maximum-circulation method[46]. The resulting input shapes were 7 × 72 × 72 for surface variables and 5 × 13 × 72 × 72 for upper-air variables.

**Model Architecture**

TIFNet implements a transformer-based encoder–decoder framework that seamlessly integrates historical cyclone dynamics with anticipated environmental drivers (Fig.1).

**Patch Embedding & Positional Encoding.** Each time-step's upper-air and surface inputs are passed through Vision Transformer[47]–style patch embedding into 324 nonoverlapping patches (each with 512 features). These are concatenated into a tensor of shape T × 324 × 1024, prefacing a learnable global token for holistic context. Sinusoidal temporal encodings[32] and learnable spatial embeddings (from TimeSformer[48]) further enrich the representation, enabling robust spatiotemporal comprehension.

**Encoder (Historical Context).** The encoder processes 24 h of pre-initialization data sampled at 6-h intervals ($T_{enc}$ = 5 time steps), encompassing multilevel atmospheric variables and the storm's historical intensity. It is designed to learn rich spatiotemporal representations that encapsulate the cyclone's recent evolution patterns.

**Decoder (Forecast Context).** The decoder ingests 120 h of post-initialization forecasted environmental data ($T_{dec}$ = 21 time steps), extracting predictive features relevant to impending intensity fluctuations.

**Divided Space–Time Attention[48].** To efficiently capture spatiotemporal features, the model employs a multi-head divided space-time attention. Temporal attention $\mathcal{A}_{\text{time}}$ is first applied independently at each spatial location to enhance temporal correlations. Spatial attention $\mathcal{A}_{\text{space}}$ is then applied at each time step to model spatial dependencies. This two-stage attention mechanism significantly improves both computational efficiency and the model's ability to represent spatiotemporal features.

$$\text{Temporal Attention: } \tilde{X}_{:,n} = \mathcal{A}_{\text{time}}(X_{:,n}), n = 1, \ldots, 325 \tag{1}$$

$$\text{Spatial Attention: } \hat{X}_{t,:} = \mathcal{A}_{\text{space}}(\tilde{X}_{t,:}), t = 1, \ldots, T, T \in \{T_{\text{enc}}, T_{\text{dec}}\} \tag{2}$$

**Cross-Attention and Prediction.** To align history with forecast, we apply a cross-attention mechanism: decoder outputs (queries) attend to encoder outputs (keys and values), blending information from both temporal domains. The fused features Z pass through a multi-head self-attention mechanism to capture dependencies across future lead times. A shared feedforward neural network (FNN) then predicts each lead's intensity change, $\Delta Vmax$ which then added to the initial intensity ($\Delta Vmax_{t_0}$), yields the full forecast trajectory.

$$Q = Z_{\text{dec}} W_Q, K = Z_{\text{enc}} W_K, V = Z_{\text{enc}} W_V \tag{3}$$

$$Z = \text{Softmax}\left(\frac{QK^\top}{\sqrt{1024}}\right) V \tag{4}$$

$$\Delta Vmax = \text{FNN}(Z) \in \mathbb{R}^{T_{\text{dec}}} \tag{5}$$

**Training Strategy**

Real-time typhoon intensity forecasting requires a model trained on operational data. However, the historical ECMWF analysis and forecast archives are both limited in terms of length and non-stationary, owing to frequent upgrades in the forecasting system. This challenge is common across large-scale weather models, where the requirements for long, consistent datasets often conflict with the evolving nature of operational systems. A widely adopted solution—also used here—is a two-stage training strategy that balances data volume and consistency.

**Stage 1: Pretraining with Reanalysis Data.** We first pretrained TIFNet using ERA5 data for both pre- and post-initialization inputs. This approach leverages dynamically consistent and unbiased data, providing stable physical constraints that help shape robust spatiotemporal representations.

**Stage 2: Fine-Tuning with Forecast Data.** Next, we fine-tuned only the final FNN using ECMWF analysis (pre-initialization) and forecast (post-initialization) fields. This stage exposes the model to realistic forecast uncertainties and addresses spatial resolution and bias differences (e.g., 0.125° vs. 0.25°). By freezing the encoder–decoder backbone, we preserve the high-level structures learned during pretraining while adapting to operational data shifts.

**Optimization Setup.** Training was conducted on a single NVIDIA A6000 GPU. We used the mean squared error as the loss function, with a batch size of 8 and weight decay of $1 \times 10^{-3}$. The Adam optimizer (momentum = 0.9)[49] was combined with a cosine annealing learning rate schedule, featuring a 3-epoch warm-up, a 10-epoch cycle, a decay factor of 0.8, and a learning rate range from $1 \times 10^{-6}$ to $4 \times 10^{-4}$ ($1 \times 10^{-6}$ to $1 \times 10^{-5}$ for the fine-tuning stage). Both pretraining and fine-tuning ran for up to 20 epochs, with early stopping triggered after 5 epochs without validation improvement.

**DATA AVAILABILITY**

The ERA5 dataset is accessible through the Copernicus Climate Data Store (CDS). The forecast data from the ECMWF IFS, the NCEP GFS and the CMA TYM were obtained from CMA's internal website. Publicly available ECMWF IFS data can be accessed through the official TIGGE archive at https://confluence.ecmwf.int/display/TIGGE. Publicly available NCEP GFS data can be accessed at https://rda.ucar.edu/datasets/d084001.

**ACKNOWLEDGMENTS**

This study was supported by the National Key R&D Program of China under grant 2023YFC3008005, the National Natural Science Foundation of China under grants 42375015, and 42192554, the Typhoon Scientific and Technological Innovation Group of the China Meteorological Administration under grant CMA2023ZD06, and the Basic Research Fund of CAMS under grant 2023Z020, S&T Development Fund of CAMS under Grant 2024KJ018 and 2024KJ022.

**AUTHOR CONTRIBUTIONS**

H.Q. and H.X. designed the research; H.Q. performed the model training and evaluation; H.Q. and H.X. wrote the original draft; H.X. conducted the review and editing; and all the authors contributed to the interpretation of the results and writing of the manuscript.

**COMPETING INTERESTS**

The authors have declared no conflicts of interest regarding this article.

Appendix

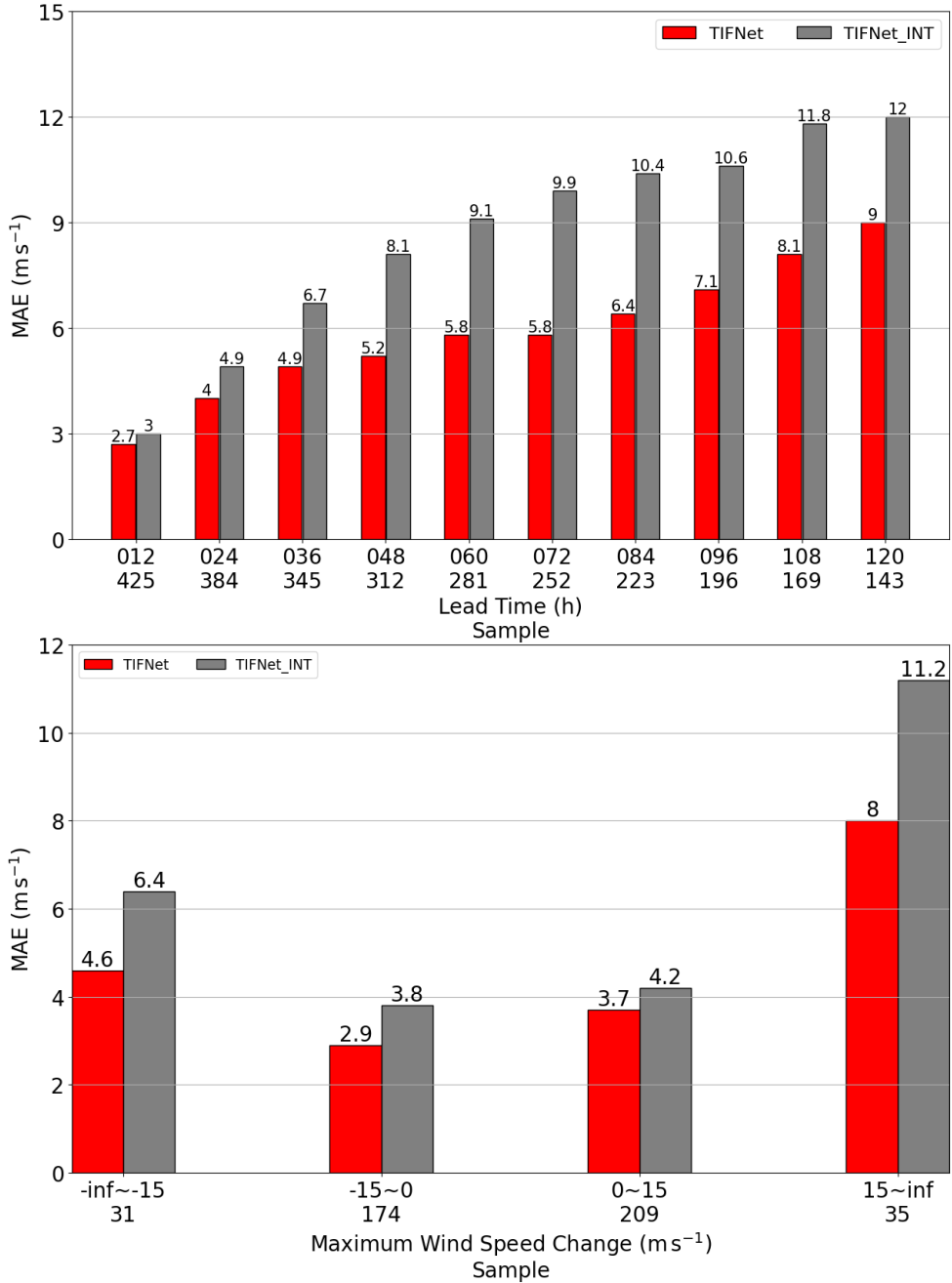

**Extended Data Fig. 1 | Impact of future environmental inputs: comparison between forecast-augmented and analysis-only TIFNet configurations (TIFNet_INT).**

**a**, Mean absolute error (MAE; m s$^{-1}$) of 12–120-h intensity forecasts for the standard TIFNet configuration (red), which incorporates IFS forecast fields from 0-120 h together with historical intensity records, and a variant (TIFNet_INT; gray) that uses only IFS analysis fields at initialization (0 h) with no future forecast input.

**b**, Forecast skill stratified by intensity-change regime, showing the MAE (m s$^{-1}$) for TIFNet (red) and TIFNet_INT (gray) across 24-h intensity change bins ($\Delta Vmax_{24}$, m s$^{-1}$): $\leqslant -15$, $-15$ to $0$, $0$ to $15$, and $\geqslant 15$. MAE values are shown above each bar; sample sizes are indicated below the x-axis.

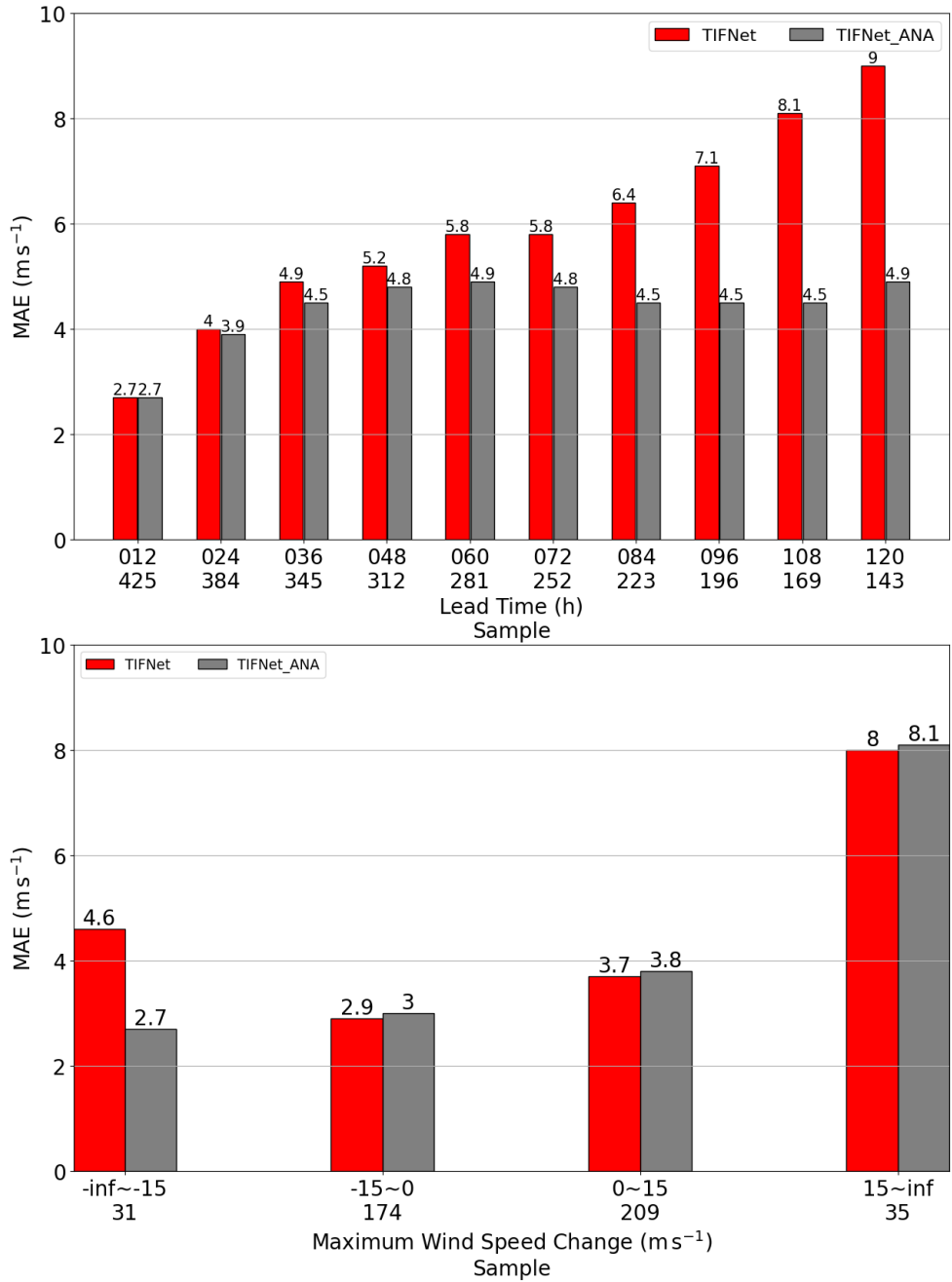

**Extended Data Fig. 2 | Impact of forecast inputs: comparison between TIFNet and the analysis-driven variant (TIFNet_ANA).**

**a**, Mean absolute error (MAE; m s$^{-1}$) of 12–120-h intensity forecasts for the standard TIFNet configuration (red), which uses IFS forecast fields from 0-120 h together with historical intensity records, and a variant (TIFNet_ANA; gray) in which all IFS forecast inputs were replaced with corresponding IFS analysis fields at each forecast hour.

**b**, MAE (m s$^{-1}$) across intensity-change regimes for TIFNet (red) and TIFNet_ANA (gray), stratified by $\Delta Vmax_{24}$ as in Extended Data Fig. 1. MAE values are shown above each bar; sample sizes are indicated below the x-axis.

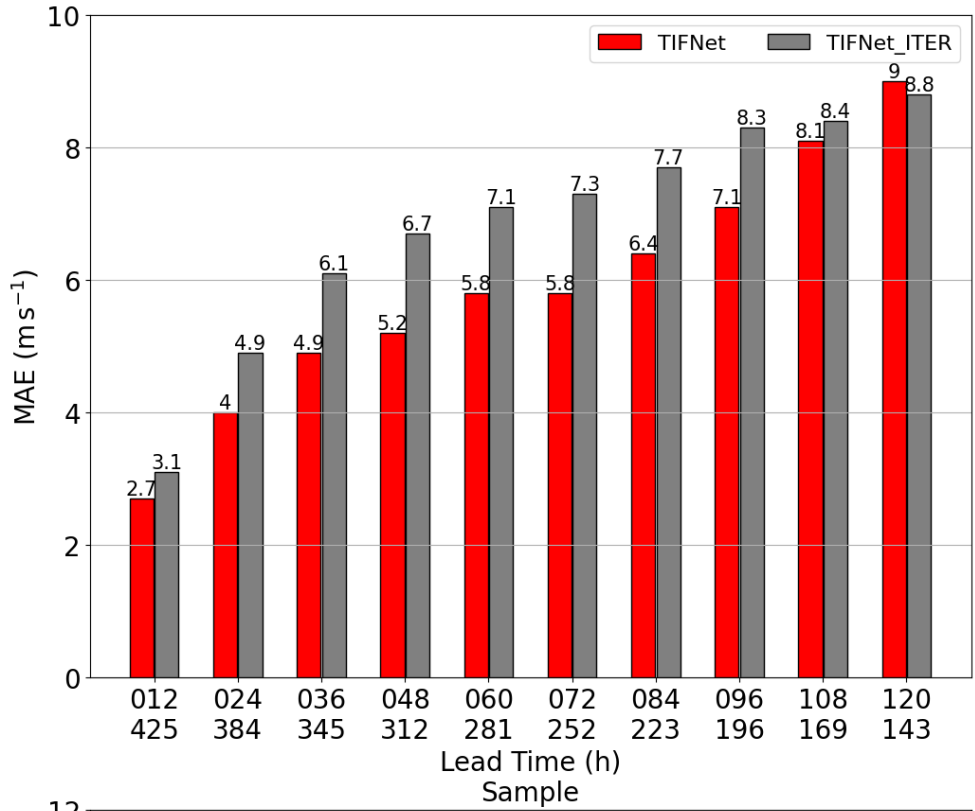
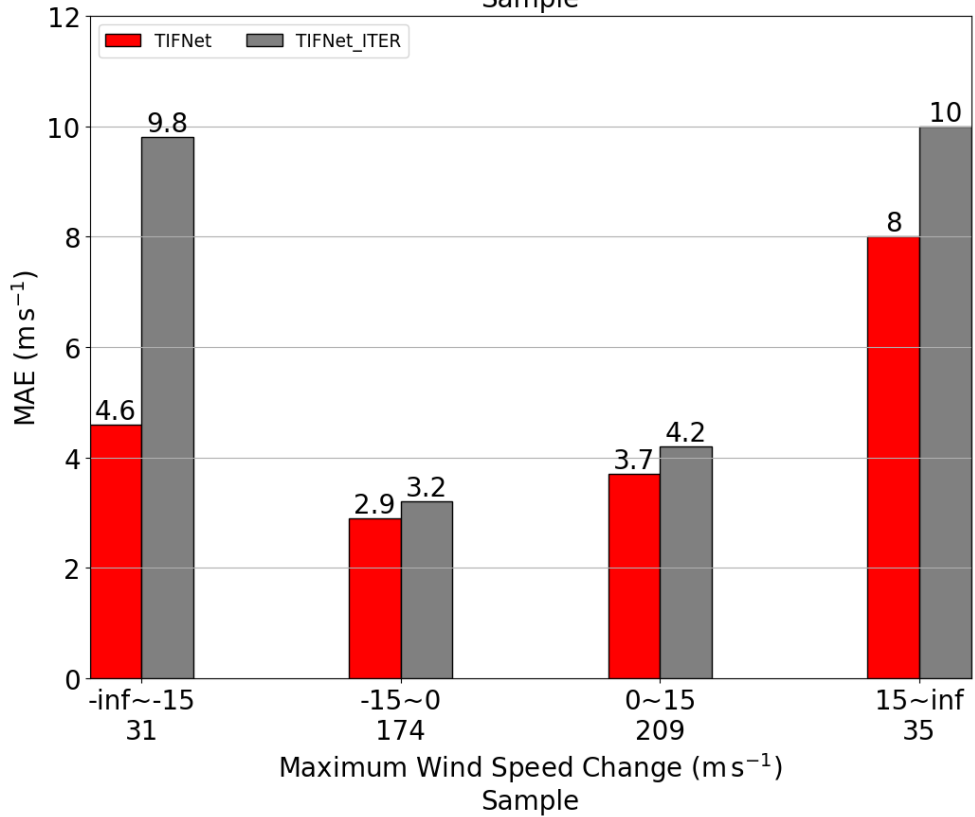

**Extended Data Fig. 3 | Forecast skill comparison between non-iterative and iterative TIFNet**

**variants.**

**a**, MAE (m s$^{-1}$) of 12–120-h intensity forecasts for standard TIFNet (non-iterative; red) and an iterative version (TIFNet_ITER; gray).

**b**, MAE (m s$^{-1}$) across intensity-change regimes for TIFNet (red) and TIFNet_ITER (gray), stratified by $\Delta Vmax_{24}$ as in Extended Data Fig. 1. MAE values are shown above each bar; sample sizes are indicated below the x-axis.

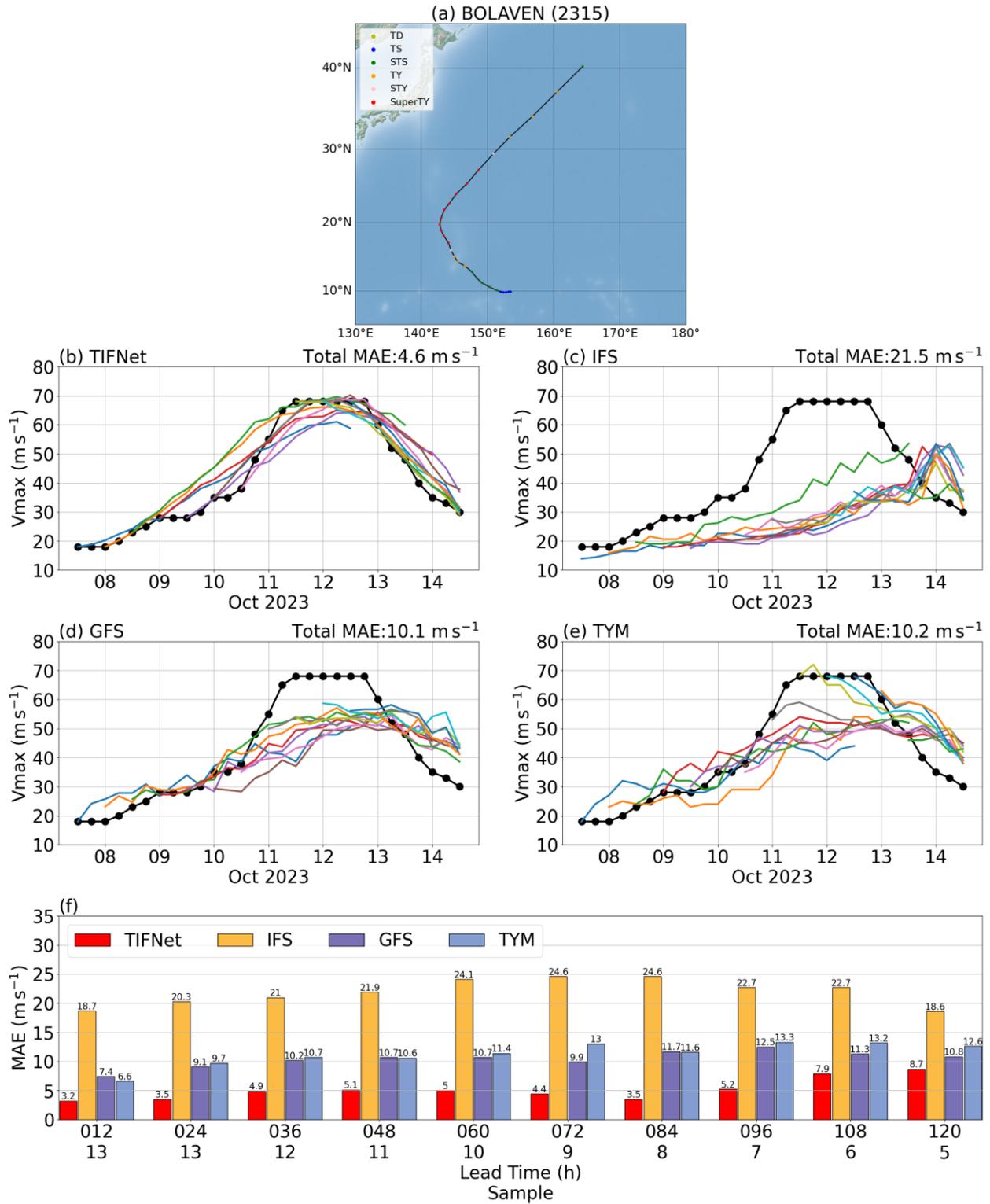

**Extended Data Fig. 4 | Forecast validation for Typhoon BOLAVEN (2315).**

(a) Tropical cyclone tracks from the CMA best track, color-coded by intensity category. (b–e) Cycle

intensity forecasts from TIFNet, the ECMWF IFS, the NCEP GFS, and the CMA TYM compared with best-track observations (black), with different colors denoting forecasts from successive initialization times. Reported values indicate the total mean absolute error (MAE; m s$^{-1}$) for each model. (f) Lead-time-dependent MAE (12–120 h) for the four models, with sample sizes shown below the x-axis.

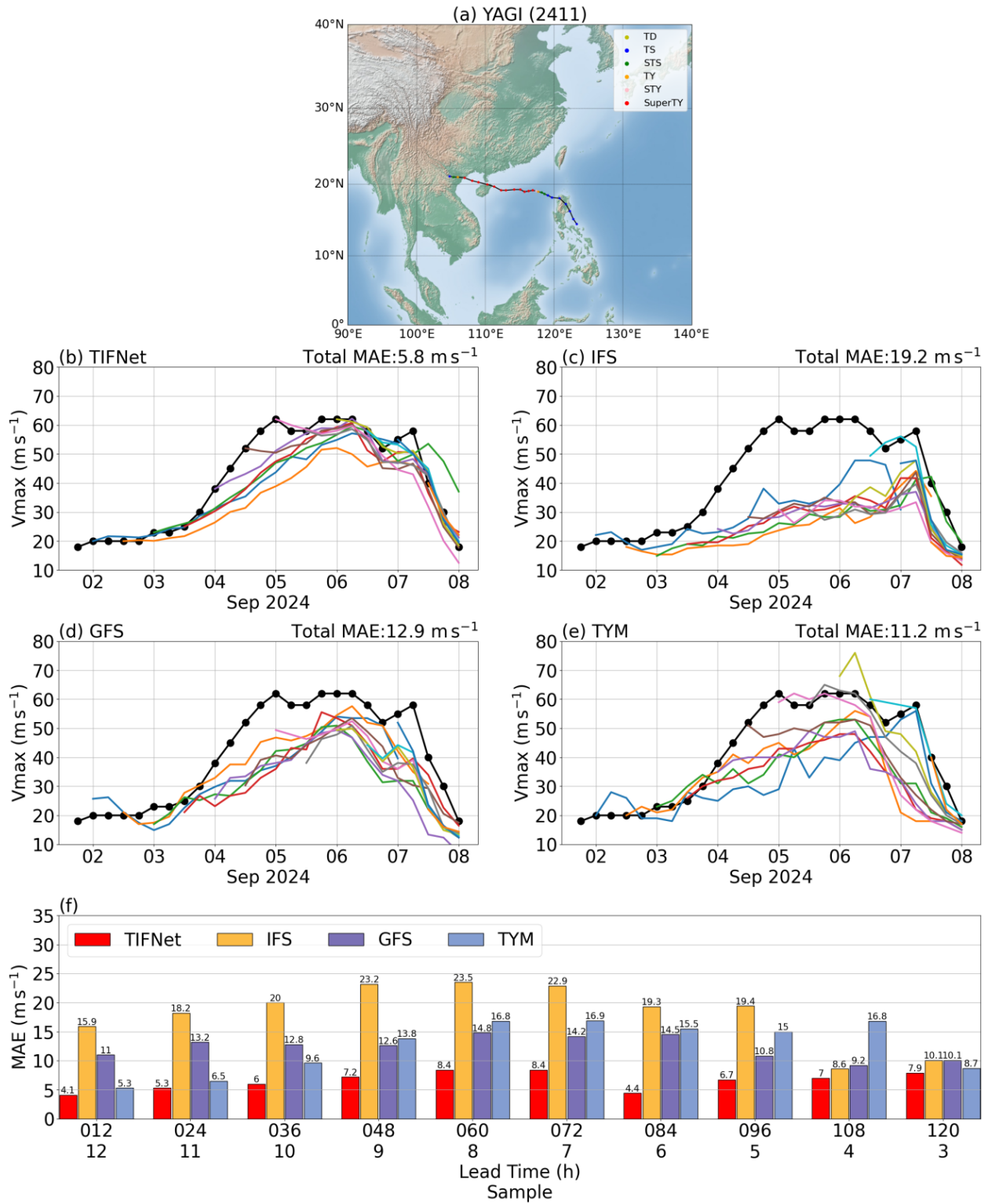

**Extended Data Fig. 5 | Forecast validation for Typhoon YAGI (2411).**

(a) Tropical cyclone tracks from the CMA best track, color-coded by intensity category. (b–e) Cycle

intensity forecasts from TIFNet, the ECMWF IFS, the NCEP GFS, and the CMA TYM compared with best-track observations (black), with different colors denoting forecasts from successive initialization times. Reported values indicate the total mean absolute error (MAE; m s$^{-1}$) for each model. (f) Lead-time-dependent MAE (12–120 h) for the four models, with sample sizes shown below the x-axis.

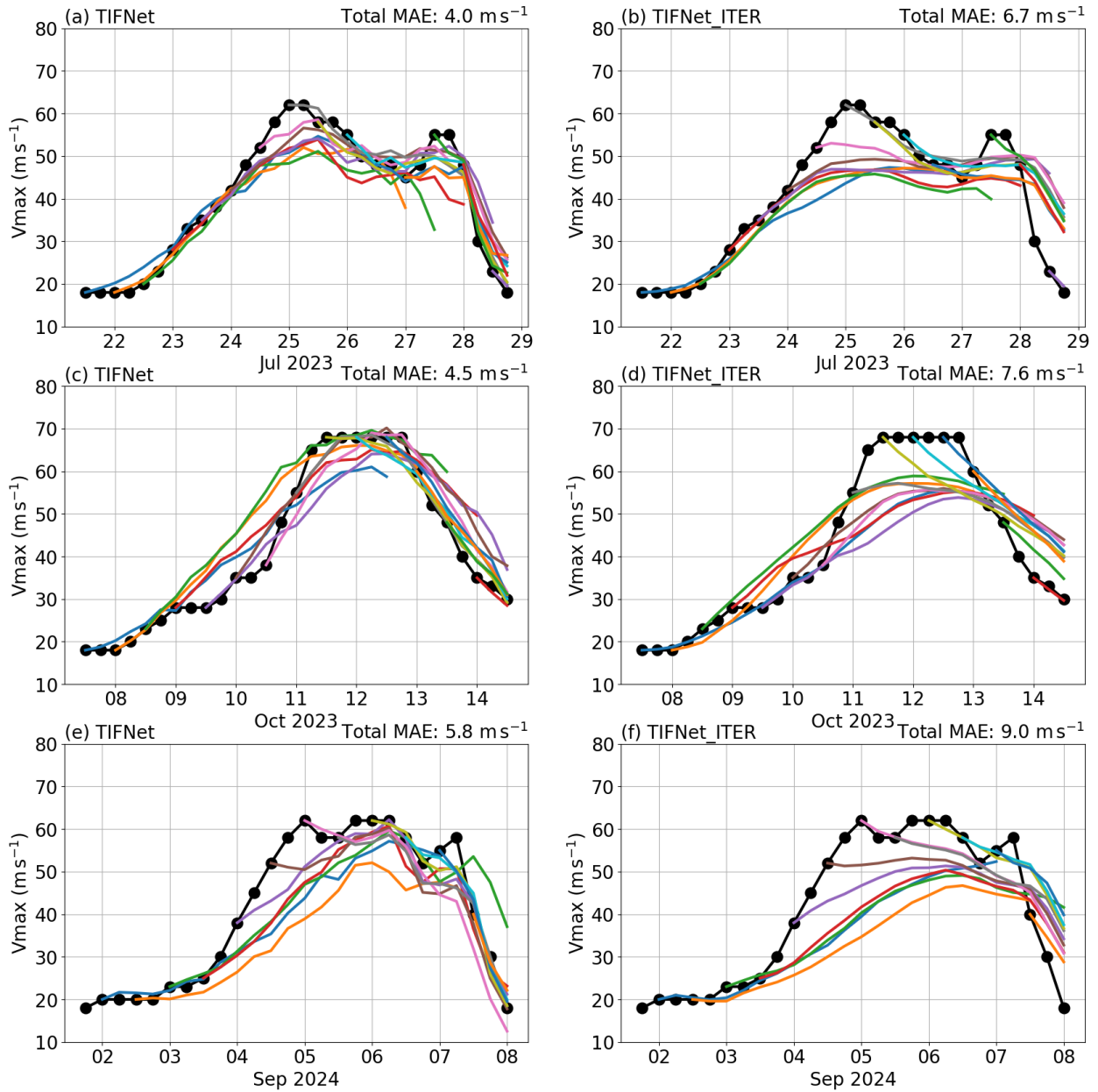

**Extended Data Fig. 6 | Forecast validation for typhoons DOKSURI, BOLAVEN, and YAGI, comparing TIFNet and TIFNet_ITER.**

(a) Intensity forecasts from the standard non-iterative TIFNet for Typhoon DOKSURI (2305), overlaid with best-track observations (black). (b) Forecasts from the iterative variant TIFNet_ITER for the same event. (c), (d) Same as (a) and (b), respectively, but for Typhoon BOLAVEN (2315). (e), (f) Same as (a) and (b), respectively, but for Typhoon YAGI (2411). Across all three cases, the non-iterative TIFNet better captures rapid intensity changes and peak magnitudes, while the iterative version produces smoother trajectories with delayed responses.